\documentclass[11pt]{article}
\usepackage{amsmath,amssymb,epsfig,bm}

\date{\today}

\begin{document}

\date{\empty}

\title{\textbf{Kinematical and dynamical aspects of ghost-matter cosmologies}}

\author{Ameya Chavda${}^1$, John D. Barrow${}^2$ and Christos G. Tsagas${}^3$\\ {\small ${}^1$Department of Physics, Columbia University, New York, NY 10027, USA}\\ {\small $^2$DAMTP, Centre for Mathematical Sciences, University of Cambridge}\\ {\small Wilberforce Road, Cambridge CB3 0WA, UK}\\ {\small ${}^3$Section of Astrophysics, Astronomy and Mechanics, Department of Physics}\\ {\small Aristotle University of Thessaloniki, Thessaloniki 54124, Greece}}

\maketitle

\begin{abstract}
We consider the kinematical and dynamical evolution of Friedmann universes with a mixture of non-interacting matter and a ghost-like field, in a scenario analogous to that advocated by the Quintom model. Assuming that the conventional matter dominates today, we find that the ghost component can bring the future expansion and the past contraction of the model to a finite halt. Moreover, at the moment the expansion or contraction stops, we find that the tendency of the universe is to bounce back and re-collapse or re-expand. Therefore, the presence of a (never dominant) ghost-field with negative density could, in principle, drive the universe into an eternal cycle of finite expansion, collapse, and re-expansion. Our study outlines the key features of such a scenario and provides a simple condition for it to occur. We also derive an autonomous set of differential equations and then employ dynamical-system techniques to identify two families of fixed points, with and without spatial curvature respectively. The members of the first family correspond to coasting universes and are stable in the Lyapunov sense. Those of the latter family are unstable repellers when their matter satisfies the strong energy condition and Lyapunov stable in the opposite case.
\end{abstract}

\section{Introduction}\label{sI}
Ghost-like fields, namely matter with negative energy density have a fairly long (though discontinuous) research history in cosmology, which can be traced back to the Steady-State theory~\cite{HN}. The idea resurfaced around the turn of the millennium~\cite{C,CKW}, with the advent of the accelerating universe and the scenarios proposed to explain it (see~\cite{G} for an interesting discussion of the subject). In addition to the dark-energy question, ghost-fields have also been employed to address phenomenological issues such as inflation and the cyclic-universe models. In particular, ghosts appear inevitable in bouncing and cyclic cosmologies (e.g.~see~\cite{ST1}-\cite{ACG}), where it is the negative energy contribution of the ghost that forces the Hubble parameter to zero. Theoretically, ghost-type matter can be realised in more than one way and the typical example is to introduce a scalar field with a Lagrangian that allows its (effective) energy density to turn negative. (e.g.~see~\cite{FFKL}-\cite{MTT} and references
therein). Ghost-like behaviour, without explicitly involving ghosts is also possible (e.g.~see~\cite{SZ,VKS}). Here, our aim is to study the kinematical and dynamical phenomenology of Friedmann-Robertson-Walker (FRW) cosmologies containing a non-interacting mixture of ordinary and ghost-like matter. In this respect, our scenario is similar to the more fimiliar``Quintom'' model, which also allows for a conventional and a ghost scalar field, with canonical and negative kinetic terms respectively (e.g.~see~\cite{Cetal1}-\cite{CPRS}). Here, however, the ghost component never dominates over its conventional counterpart and the total energy density of the universe remains non-negative at all times. Moreover, the effective equation of state is not allowed to enter the ``tachyon'' or the ``phantom'' regime.\footnote{Ghosts have been reported to suffer from instabilities at the quantum level (unbounded vacuum decay -- e.g.~see~\cite{CHT,KKLM}). The debate is still going on, however, with other studies claiming that such instabilities could be alleviated in time-dependent cosmological backgrounds (e.g.~see~\cite{ALCS}-\cite{SS}). It might also turn out that we are dealing with an effective theory~\cite{CJM}. As a result, studies of ghost, or Quintom-like, models are fairly popular in the literature. Here, we will not enter the discussion of the aforementioned quantum-instability problem, which may only be settled by a successful future modification of gravity. Instead, we will consider the classical phenomenology of FRW-like cosmological models containing a sub-dominant ghost component.} These mark some of the differences between our study and other analogous phenomenological treatments, such as those of~\cite{MG}-\cite{MTT} for example. Proceeding as in~\cite{BKM,BT}, we start from a Lagrangian which guarantees negative kinetic energy for the associated scalar field. The latter has a stress-energy tensor that corresponds to a perfect fluid and it is therefore compatible with the symmetries of the FRW spacetimes. This, in turn, enables us to investigate the implications of a ghost field  for the kinematics and the dynamics of the standard cosmological model. We note that there has been recent interest in cosmologies that undergo a change in the sign of their effective density at late times due to a change of sign in the cosmological 'constant' contribution. Detailed analyses show that these models have favorable consequences for the Hubble-tension problem~\cite{SSS}-\cite{ABEA}.

Given its negative contribution to the total energy density of the FRW universe, the effect of the ghost component resembles that of a positive cosmological constant. More specifically, adopting a flat Friedmann model and assuming that the ordinary matter dominates completely at present, we show that the ghost fluid will bring the expansion to a halt within a finite time in the future, or ``remove'' the singularity from the universe's past (or perhaps even both). In so doing, we also provide the condition leading to the former, or to the latter, of these scenarios. Put another way, the presence of the ghost field guarantees that there is a static moment in the future, or in the past, of the FRW host (or both). Interestingly, when the static moment occurs in the future, the tendency of the universe is to re-collapse, whereas the past static moment is followed by a bounce and re-expansion.\footnote{Certain aspects of the scenarios presented here have also been discussed in~\cite{MG,MTT}, although there the adopted scalar fields had different physical characteristics (see \S~\ref{ssBCMs} below).} Therefore, in principle at least, one could have a simple ``cyclic'' cosmological model that expands, contracts and re-expands eternally, while it periodically goes through phases of accelerated expansion as well (see \S~\ref{ssCMK=0} and \S~\ref{ssBCMs}). It also important to note that all these epochs have finite duration (i.e.~there are no past or future singularities) and in none of them the ghost component needs to dominate over the ordinary matter. In particular, the ghost-like field can be arbitrarily weak today.

We then turn our attention to the (dynamical) stability of FRW cosmologies with a mixture of ordinary and ghost matter. Employing dynamical system techniques and demanding that both of the aforementioned fields are present at all times, we distinguish between two families of equilibrium states (fixed points). These are classified depending on whether the spatial curvature of the host spacetime vanishes or not. We find that both sets of fixed points correspond to non-static cosmologies, which are rather severely constrained and therefore offer limited phenomenology.\footnote{The linear stability of static Friedmann universes with a mixture of ghost-like and ordinary matter has been investigated in~\cite{BT}, using relativistic cosmological perturbation theory (see also \S~\ref{ssSESs} here).} Allowing for nonzero 3-curvature, in particular, leads to equilibrium states that correspond to ``coasting'' Friedmann universes. These turn out to be linearly stable, though not asymptotically but only in the Lyapunov sense. When the geometry of the spatial sections is Euclidean, on the other hand, the associated $3\times3$ dynamical system accepts one zero eigenvalue and the stability issue becomes less straightforward to deduce. In order to bypass the zero-eigenvalue obstacle, we look for specific analytic solutions. The later show instability when the ordinary and the ghost-like matter fields satisfy the strong energy condition and Lyapunov-type stability in the opposite case.

\section{Fluid description of ghost scalar fields}\label{sFDGSFs}
Scalar fields can be treated as effective fluids with an equation of state that depends on the balance between their kinetic and potential energies. Minimally coupled scalar fields, in particular, are known to correspond to effective perfect fluids (e.g.~see~\cite{M}).

\subsection{$\psi$-field stress-energy tensor}\label{sspsiSET}
Consider a general spacetime with arbitrary metric $g_{ab}=g_{ab}(x^c)$, containing a minimally coupled scalar field ($\psi=\psi(x^a)$) with Lagrangian
\begin{equation}
\mathcal{L}_{\psi}= \sqrt{-g}\left[{1\over2}\,\nabla_a\psi\nabla^a\psi+V(\psi)\right]\,,  \label{Lpsi}
\end{equation}
where $V=V(\psi)$ is the potential of the $\psi$-field. The above Lagrangian corresponds to a stress-energy tensor of the form (e.g.~see~\cite{TCM})
\begin{equation}
T_{ab}^{(\psi)}= -\nabla_a\psi\nabla_b\psi +\left[{1\over2}\,\nabla_c\psi\nabla^c\psi+V(\psi)\right]g_{ab}\,.  \label{Tpsi1}
\end{equation}
Assuming that $\nabla_a\psi\neq0$, the familiar conservation law $\nabla^bT^{(\psi)}_{ab}=0$ leads to the associated Klein-Gordon equation, namely
\begin{equation}
\nabla^a\nabla_a\psi- V^{\prime}(\psi)= 0\,,  \label{KGpsi1}
\end{equation}
with the prime denoting differentiation with respect to $\psi$. Note that when $\nabla_a\psi=0$, expression (\ref{Tpsi1}) reduces to $T_{ab}^{(\psi)}=V(\psi)g_{ab}$. This and the energy-momentum conservation imply that $\nabla_aV(\psi)=0$, which ensures that $\psi$ acts as an effective cosmological constant rather than a dynamical scalar field.

\subsection{$\psi$-field kinematics}\label{sspsiFKs}
Let us assign a 4-velocity vector to the $\psi$-field. In particular, demanding that $\nabla_a\psi\nabla^a\psi<0$ over an open spacetime region, the gradient $\nabla_a\psi$ defines the timelike 4-velocity field
\begin{equation}
u_a= -{1\over\dot{\psi}}\,\nabla_a\psi\,,  \label{ua}
\end{equation}
where $\dot{\psi}=u^a\nabla_a\psi\neq0$ by construction (e.g.~see \S~2.6.2 in~\cite{TCM} for the technical details). In addition, $\dot{\psi}^2= -\nabla_a\psi\nabla^a\psi>0$ and $u_au^a=-1$ as required. The $u_a$-field also defines the direction of time and introduces an 1+3 splitting of the spacetime into time and 3-space. The metric of the latter is given by the projector
\begin{equation}
h_{ab}= g_{ab}+ {1\over\dot{\psi}^2}\nabla_a\psi\nabla_b\psi\,,  \label{hab}
\end{equation}
which satisfies the constraints $h_{ab}=h_{ba}$, $h_{ab}u^b=0$, $h_a{}^a=3$ and $h_{ac}h^c{}_b= h_{ab}$. Moreover, the projection tensor defines the covariant derivative operator (${\rm D}_a=h_a{}^b\nabla_b$) in the 3-D hypersurface orthogonal to the $u_a$-field. It is then straightforward to show that ${\rm D}_a\psi=0$~\cite{TCM}.

The full kinematical description of the $\psi$-field follows by decomposing the gradient of the associated 4-velocity vector as (e.g.~see~\cite{TCM})
\begin{equation}
\nabla_bu_a= {1\over3}\,\Theta h_{ab}+ \sigma_{ab}+ \omega_{ab}- A_au_b\,,  \label{Nbua}
\end{equation}
where $\Theta=\nabla^au_a={\rm D}^au_a$ is the expansion/contaction scalar, $\sigma_{ab}={\rm D}_{\langle b}u_{a\rangle}$ is the shear tensor, $\omega_{ab}={\rm D}_{[b}u_{a]}$ is the vorticity tensor and $A_a=\dot{u}_a$ is the 4-acceleration vector.\footnote{Round brackets indicate symmetrisation, square ones antisymmetrisation and angled brackets denote the symmetric traceless component of second rank tensors. For instance, ${\rm D}_{\langle b}u_{a\rangle}= {\rm D}_{(b}u_{a)}-({\rm D}^cu_c/3)h_{ab}$.} Thus, employing definition (\ref{ua}), we find that in our case
\begin{equation}
\Theta= -{1\over\dot{\psi}}\left(\ddot{\psi}+V^{\prime}\right)\,, \hspace{10mm} \sigma_{ab}= {1\over3\dot{\psi}} \left(\ddot{\psi}+V^{\prime}\right)h_{ab} -{1\over\dot{\psi}}\,{\rm D}_a{\rm D}_b\psi\,, \hspace{10mm} \omega_{ab}= 0  \label{psiKin1}
\end{equation}
and
\begin{equation}
A_a= -{1\over\dot{\psi}}\,{\rm D}_a\dot{\psi}\,,  \label{psiKin2}
\end{equation}
with ${\rm D}_a{\rm D}_b\psi=h_a{}^ch_b{}^d\nabla_c\nabla_d\psi$  by construction~\cite{TCM}. Consequently, the $\psi$-field is irrotational, it has nonzero shear and it can generally expand or contract (when $\Theta\gtrless0$ respectively). In addition, following Eq.~(\ref{psiKin2}), $\dot{\psi}$ acts as an effective 4-acceleration potential.

\subsection{$\psi$-fields as ghost matter}\label{sspsiFGM}
Introducing the timelike $u_a$-field also facilitates a fluid-description for the $\psi$-field. More specifically, on using definition (\ref{ua}), the energy-momentum tensor given in Eq.~(\ref{Tpsi1}) assumes the perfect-fluid form
\begin{equation}
T_{ab}^{(\psi)}= \rho^{(\psi)}u_au_b+ p^{(\psi)}h_{ab}\,,  \label{Tpsi2}
\end{equation}
where
\begin{equation}
\rho^{(\psi)}= -{1\over2}\,\dot{\psi}^2- V \hspace{15mm} {\rm and} \hspace{15mm} p^{(\psi)}= -{1\over2}\,\dot{\psi}^2+ V\,,  \label{pfpsi}
\end{equation}
represent the effective energy density and isotropic pressure of the $\psi$-field. Therefore, as long as $\dot{\psi}^2+2V>0$, the energy density is negative definite and we are dealing with a ghost-like scalar field. When $|V|\ll\dot{\psi}^2$, expressions (\ref{pfpsi}) reduce to  $p^{(\psi)}\simeq\rho^{(\psi)}\simeq -\dot{\psi}^2/2<0$, which corresponds to ``stiff'' ghost matter (e.g.~see \cite{BT}). Alternatively, that is for $\dot{\psi}^2\ll|V|$, we have $p^{(\psi)}\simeq-\rho^{(\psi)}\simeq V$. Note that in both of the aforementioned special cases the associated matter field behaves as a barotropic medium with $p^{(\psi)}\simeq p^{(\psi)}(\rho^{(\psi)})$. In general, however, we have $p^{(\psi)}=\rho^{(\psi)}+2V=-\rho^{(\psi)}-\dot{\psi}^2$ and the effective equation of state of the matter field described by (\ref{pfpsi}) is determined by the $w^{(\psi)}$-parameter
\begin{equation}
w^{(\psi)}= {p^{(\psi)}\over\rho^{(\psi)}}= {\dot{\psi}^2-2V\over\dot{\psi}^2+2V}\,,  \label{wpsi}
\end{equation}
so that $-1<w^{(\psi)}<1$. Given that $\dot{\psi}^2>0$ and assuming ghost-like matter (with $\dot{\psi}^2+2V>0$ at all times -- see Eq.~(\ref{pfpsi}a) above), the lower bound on $w^{(\psi)}$ holds always. The upper bound, on the other hand, applies as long as $V>0$. In the opposite case, namely when $V<0$ (with $\dot{\psi}^2+2V>0$ always), we may have $w^{(\psi)}>1$. Such models (with an effective sound speed grater than that of light) have been explored in the past, particularly in the context of the so-called ``cyclic universes'' (e.g.~see~\cite{EWST}-\cite{LBL}). On the other hand, our earlier assumption of a ghost-like fluid, excludes the possibility of ``phantom matter'' with $w<-1$.

Finally, it is worth mentioning that, once the timelike 4-velocity field (\ref{ua}) has been introduced, we have $\nabla_a\psi= -\dot{\psi}u_a$ (recall that ${\rm D}_a\psi=0$). Then, the Klein-Gordon equation (see expression (\ref{KGpsi1}) earlier) recasts into
\begin{equation}
\ddot{\psi}+ \Theta\dot{\psi}+ V^{\prime}= 0\,.  \label{KGpsi2}
\end{equation}
Note that this relation also provides the conservation law of the ghost energy density. Indeed, on using (\ref{pfpsi}), the above form of the Klein-Gordon equation becomes
\begin{equation}
\dot{\rho}^{(\psi)}= -\Theta(\rho^{(\psi)}+p^{(\psi)})\,,  \label{psiedcl}
\end{equation}
which is the continuity equation of the ghost matter.

\section{Friedmann universes with ghost and regular 
matter}\label{sFUGRM}
The perfect fluid description of the ghost-like scalar field makes it compatible with the high symmetry of the FRW spacetimes. In what follows, we will examine the implications of such a ghost-matter component for the evolution of typical Friedmann universes.

\subsection{Equations and key features}\label{ssEKFs}
Consider an FRW cosmology, without a cosmological constant, containing regular matter and a non-interacting ghost component. The evolution of the aforementioned model is governed by the system
\begin{equation}
H^2= {1\over3}\,\kappa\rho+ {1\over3}\,\kappa\rho^{(\psi)}- {K\over a^2}\,,  \label{Fried1}
\end{equation}
\begin{equation}
\dot{H}= -H^2- {1\over6}\,\kappa\rho(1+3w)- {1\over6}\,\kappa\rho^{(\psi)}(1+3w^{(\psi)})\,,  \label{Ray1}
\end{equation}
\begin{equation}
\dot{\rho}= -3H(1+w)\rho  \label{mcont1}
\end{equation}
and
\begin{equation}
\dot{\rho}^{(\psi)}= -3H(1+w^{(\psi)})\rho^{(\psi)}\,,  \label{gcont1}
\end{equation}
where $\kappa=8\pi G$ and $c=1$. These are respectively the Friedmann, the Raychaudhuri and the continuity equations of the two matter fields. Also, $H=\dot{a}/a$ is the Hubble parameter, with $a=a(t)$ representing the cosmological scale factor, $\rho=\rho(t)$ is the energy density of the conventional matter, $p=p(t)$ is its isotropic pressure and $w=p/\rho$ is the associated barotropic index. Finally, $K=0,\pm1$ is the 3-curvature index, with $K=0$ corresponding to a flat FRW cosmology and $K=\pm1$ indicating a closed/open universe respectively.

The contribution of the matter fields to the total energy density of the system is measured by the associated density parameters, which in our case are $\Omega_{\rho}=\kappa\rho/3H^2>0$ for the conventional matter and $\Omega_{\psi}=\kappa\rho^{(\psi)}/3H^2<0$ for the ghost component. Similarly, the parameter $\Omega_K=-K/a^2H^2$ measures the (effective) curvature input to the energy density of the model. Employing the above, the Friedmann equation reads
\begin{equation}
1= \Omega_{\rho}+ \Omega_{\psi}+ \Omega_K\,.  \label{Fried2}
\end{equation}
Given that $\Omega_{\psi}<0$ and $\Omega_K=-K/a^2H^2$, the above implies that the ghost component could play (to some extent at least) the role of positive spatial curvature. Also, recalling that $q=-\ddot{a}a/\dot{a}^2=-[1+(\dot{H}/H^2)]$ defines the deceleration parameter of the universe, we may rewrite Raychaudhuri's formula as
\begin{equation}
q= {1\over2}\, \left[\Omega_{\rho}+\Omega_{\psi}+3(w\Omega_{\rho} +w^{(\psi)}\Omega_{\psi})\right]\,.  \label{q1}
\end{equation}
The latter implies that even a subdominant ghost component (i.e.~one with $|\Omega_{\psi}|<\Omega_{\rho}$) could in principle play the role of dark energy and lead to accelerated expansion. For example, when the conventional matter is in the form of pressureless dust (i.e.~for $w=0$), the deceleration parameter will take negative values if $-\Omega_{\rho}<\Omega_{\psi}< -\Omega_{\rho}/(1+3w^{(\psi)})$.

\subsection{Critical moments with $K=0$ and
$H\rightarrow0$}\label{ssCMK=0}
As we mentioned above, the fact that $\rho^{(\psi)}<0$ implies that the ghost matter can play the role of positive spatial curvature. With this in mind, let us consider a flat FRW universe with a mixture of conventional and ghost-like matter. Then, given that $H^2\geq0$ always, the Friedmann equation  (see expression (\ref{Fried1})) demands that $\rho\geq\rho^{(\psi)}$ at all times. If the conventional matter component dominates throughout the lifetime of the universe, namely when $\rho>\rho^{(\psi)}$ always, the expansion proceeds unimpeded. It is conceivable, however, that $|\rho^{(\psi)}|\rightarrow\rho$, in which case $H\rightarrow0$ (see Eq.~(\ref{Fried1}) earlier) and the expansion/contraction of the model comes to a halt. Whether such a critical moment actually occurs and when, is determined by the evolution rates of the two matter fields. Following (\ref{mcont1}) and (\ref{gcont1}), these evolve according to
\begin{equation}
\rho= \rho_0\left({a\over a_0}\right)^{-3(1+w)} \hspace{10mm} {\rm and} \hspace{10mm} \rho^{(\psi)}= \rho^{(\psi)}_0\left({a\over a_0}\right)^{-3(1+w^{(\psi)})}\,,  \label{rhos}
\end{equation}
with $\rho_0>0$ and $\rho_0^{(\psi)}<0$. Note that the zero suffix generally indicates any given time during the evolution of the universe. Here, however, we will assume that the zero suffix corresponds to the present time. Going back to Eq.~(\ref{Fried1}), we realise that the two matter fields will cancel each other's contribution out (namely $\rho\rightarrow-\rho^{(\psi)}$) when $a\rightarrow a_*$, with
\begin{equation}
a_*= a_0\left({|\rho^{(\psi)}|\over\rho}\right) _0^{1/3(w^{(\psi)}-w)}= a_0\left({|\Omega_{\psi}|\over\Omega_{\rho}}\right) _0^{1/3(w^{(\psi)}-w)}\,,  \label{a*1}
\end{equation}
where $w^{(\psi)}\neq w$.\footnote{The special case with $w^{(\psi)}=w$ is trivial phenomenologically, because then the ghost and the conventional matter fields either cancel each other's contribution to the Friedmann equation at all times, or they never do so.} At this critical moment, the Hubble parameter tends to zero (i.e.~$H\rightarrow0$) and the expansion/contraction comes to a halt. Assuming that $|\rho^{(\psi)}|/\rho= |\Omega_{\psi}|/\Omega_{\rho}<1$ today, the time that this happens depends on the barotropic indices of the two matter fields. More specifically, when $w^{(\psi)}-w>0$ the aforementioned critical moment occurs in the past (i.e.~$a_*<a_0$), while in the opposite case it happens in the future (i.e.~$a_*>a_0$). In the former scenario, our FRW model begins from a static configuration of finite ``radius''. Alternatively, that is in the latter scenario, the universe cannot grow beyond a certain ``size''.\footnote{A similar behaviour occurs in FRW cosmologies with positive spatial curvature and a vanishing ghost component (i.e.~when $K=+1$, $\rho^{(\psi)}=0$ and $\rho\neq0$). Indeed, assuming that $|\Omega_K|\ll\Omega_{\rho}$ today, the static moment of such a universe lies in our future if $w>-1/3$ and in our past when $w<-1/3$. More specifically, the radius of the static configuration (as $H\rightarrow0$ and $a\rightarrow a_*$) is given by
\begin{equation}
a_*= a_0 \left({|\Omega_K|\over\Omega_{\rho}}\right)_0^{1/[2-3(1+w)]}\,,  \label{a*2}
\end{equation}
with the zero suffix corresponding to the present. Then, when $w>-1/3$, we have $2-3(1+w)<0$ and $a_*\gg a_0$ (i.e.~future static). Alternatively, for $w<-1/3$, we find that $a_*\ll a_0$ (i.e.~past static). Recall that $\rho\propto a^{-3(1+w)}$ irrespective of the model's geometry, while the 3-curvature contribution to the Friedmann equation drops as $a^{-2}$ always (see relations (\ref{Fried1}) and (\ref{rhos}a) earlier). The close analogy in the evolution patterns of the two cosmological models should not come as a surprise, given that (qualitatively speaking at least) the ghost field mimics the contribution of positive spatial curvature to the Friedmann equation (see expressions (\ref{Fried1}) and (\ref{Fried2}) in \S~\ref{ssCMK=0}).}

One can calculate the temperature of the universe at the moment its expansion/contraction stops. Indeed, recalling that $T\propto1/a$, expression (\ref{a*1}) becomes
\begin{equation}
T_*= T_0\left({\rho\over|\rho^{(\psi)}|}\right) _0^{1/3(w^{(\psi)}-w)}= T_0\left({\Omega_{\rho}\over|\Omega_{\psi}|}\right) _0^{1/3(w^{(\psi)}-w)}\,.  \label{T*}
\end{equation}
Assuming that $H\rightarrow0$ in the past, namely that $T_*>T_0$, and given that $T_*$ cannot be lower that the temperature at nucleosynthesis (i.e.~setting $T_*>T_{BBN}$), we can use the above to put an upper bound on the energy density of any ghost component today.

\subsection{Bouncing at the critical moments}\label{ssBCMs}
At the critical moment, as the Hubble parameter approaches zero and the expansion/contraction comes to a halt, the subsequent evolution of the associated cosmological model depends on the time-derivative of the Hubble parameter. For example, when the latter also tends to zero (i.e.~if $\dot{H}\rightarrow\dot{H}_*=0$), the universe will remain static thereafter. Alternatively, there is the possibility of a bounce. The answer should come from the Raychaudhuri equation (see expression (\ref{Ray1})), which at the critical moment, namely as $\rho^{(\psi)}\rightarrow-\rho$ and $H\rightarrow0$, reduces to
\begin{equation}
\dot{H}\rightarrow \dot{H}_*= -{1\over2}\,\kappa\rho(w-w^{(\psi)})\,,  \label{Ray2}
\end{equation}
with $\rho>0$ and $w^{(\psi)}\neq w$. When $w^{(\psi)}>w$, in which case the static moment occurs in the past (see \S~\ref{ssCMK=0} previously), the above ensures that $\dot{H}_*>0$.\footnote{Recall that, when $\rho^{(\psi)}<\rho$ today, the static moment is in the past if $w^{(\psi)}>w$ and in the future for $w^{(\psi)}<w$.} This in turn ensures that our model will not reach a singularity, but instead it will bounce and start to re-expand. In the opposite case, that is when $w^{(\psi)}<w$, Eq.~(\ref{Ray2}) guarantees that $\dot{H}_*<0$. In such a case, the static moment (which now lies in the future) is followed by contraction. The latter scenario has also been discussed in~\cite{MG,MTT}, though in FRW models containing ghost-like scalar fields with a phantom-like equation of state.

Overall, a spatially flat Fridmann universe containing a mixture of conventional matter and a ghost component (with $\rho^{(\psi)}\leq\rho$ at all times and $\rho^{(\psi)}\ll\rho$ today) could have started expanding from a non-singular initial state at some finite time in the past. Alternatively, a similar model will expand for a finite period in the future before it comes to a halt and re-collapses. In either case, the universe is currently dominated by conventional matter and the timing of the bounce is determined by the relative evolution of the aforementioned two matter fields. Put another way, the presence of an arbitrarily weak ghost-like field today, can lead an FRW universe to an eternal cycle of (finite) expansion, collapse and re-expansion. Moreover, as pointed out above, the ghost component does not need to dominate over its conventional counterpart. Assuming that the barotropic index of the ordinary matter evolves between $w_i$ initially and $w_f$ at the moment of re-collapse, the aforementioned scenario will work as long as $w_i<w^{(\psi)}<w_f$.

\subsection{Critical moments with $K\neq0$ and
$H\rightarrow0$}\label{ssCKKn0}
The effects of the spatial curvature are introduced via the Friedmann equation (see (\ref{Fried1}) in \S~\ref{ssEKFs}) and their impact depends on the type of the model's geometry. In particular, when $K=+1$, the positive curvature of the 3-space acts in tune with the ghost component. In the opposite case, curvature and matter work together. We therefore expect the overall effect of the $\psi$-field to depend on its evolution relative to that of the ordinary matter and of the 3-curvature as well.

Qualitatively speaking, the evolution of an FRW universe with non-Euclidean 3-geometry and a mixture of conventional and ghost-like matter is monitored by the system of (\ref{Fried1})-(\ref{gcont1}). More specifically, employing the continuity equations of the two matter fields, one obtains (\ref{rhos}a) and (\ref{rhos}b). Substituting the latter into expression (\ref{Fried1}) leads to
\begin{equation}
H^2= {1\over3}\,\kappa\rho_0\left({a_0\over a}\right)^{3(1+w)}+ {1\over3}\,\kappa\rho^{(\psi)}_0\left({a_0\over a}\right)^{3(1+w^{(\psi)})}- {K\over a_0^2}\left({a_0\over a}\right)^2\,,  \label{Fried3}
\end{equation}
where $K=\pm1$ and the zero suffix indicates (in our case) the present time. Let us now consider a model with spherical spatial geometry and allow for matter that satisfies the strong energy condition, namely set $K=+1$ and $w>-1/3$ always. Given that the curvature term in Eq.~(\ref{Fried3}) drops like $a^{-2}$ always, the expansion will come to a halt because of the 3-curvature effects (just like in standard Friedmann cosmologies with $K=+1$), provided that $w^{(\psi)}>-1/3$. Indeed, the latter condition ensures that the $\psi$-field will never dictate the future kinematics, assuming that it is subdominant today (i.e.~that $|\Omega_{\psi}|\ll| \Omega_K|\ll\Omega_{\rho}$ at present). When $w^{(\psi)}<-1/3<w$, on the other hand, the ghost component decays slower than the curvature effects. In such a case, the evolution of the universe may reach a future static moment due to the ghost field rather than the curvature, depending on the relative values of $|\Omega_{\psi}|$, $|\Omega_K|$ and $\Omega_{\rho}$ at present. Positive curvature triggers a static moment in the past evolution of a ghost-free FRW cosmology, provided its matter violates the strong energy condition (i.e.~when $w<-1/3$ -- see footnote~6 here). Adding a $\psi$-field with $w^{(\psi)}<-1/3$ will not change the above scenario. Allowing for $w^{(\psi)}>-1/3>w$, on the other hand, could make the ghost component the reason for avoiding the initial singularity.

More dramatic are the kinematic effects of the ghost component in models with hyperbolic spatial geometry, the conventional counterparts of which are known to expand for ever. Introducing a ghost component to these models can avoid their past signularity, or bring their future expansion to a halt. The conditions for these to happen are those given in \S~\ref{ssCMK=0} earlier, plus two additional constraints due to the model's negative curvature. More specifically, when $w^{(\psi)}>w$ and $w^{(\psi)}>-1/3$ hold simultaneously, $H\rightarrow0$ at a finite time in the past. Also, the universe will not expand forever when $w^{(\psi)}<w$ and $w^{(\psi)}<-1/3$. Indeed, in the former case, the (negative) energy-density contribution of the ghost matter increases faster, than those of the conventional matter and the 3-curvature, as the universe collapses. In the latter case, on the other hand, the ghost density increases faster than the others as the universe expands. In both cases, the ghost field is assumed to be entirely subdominant at present.

Before closing this section, let us turn our attention to the Raychaudhuri equation. Evaluated at the static moment, namely as $H\rightarrow0$ and $a\rightarrow a_*$, the latter acquires an additional curvature-related term. More specifically, putting (\ref{Fried1}) and (\ref{Ray1}) together, we find that
\begin{equation}
\dot{H}\rightarrow \dot{H}_*= -{1\over6}\,\kappa\rho(w-w^{(\psi)})- {K\over2a_*^2}\,(1+3w^{(\psi)})\,.  \label{Ray3}
\end{equation}
Consequently, for $K=+1$ and $w>w^{(\psi)}>-1/3$, there is a static moment in the future, which is followed by contraction. When $w<w^{(\psi)}<-1/3$, on the other hand, there is a bounce at a finite time in the past. When dealing with negative 3-curvature, the universe stops expanding and starts to recollapse when $w^{(\psi)}<w$ and $w^{(\psi)}<-1/3$, while there is a bounce in the finite past for $w^{(\psi)}>w$ and $w^{(\psi)}>-1/3$.

\section{Equilibrium and dynamical stability}\label{sEDS}
Puting the formulae provided in \S~\ref{ssEKFs} together with the three $\Omega$-parameters defined there, we can obtain an autonomous dynamical system that describes the phase-space evolution of Friedmann universes containing a non-interacting mixture of ordinary and ghost-like matter.

\subsection{The autonomous system}\label{ssAS}
Proceeding along the lines of~\cite{UL,BTF}, we combine the Friedmann equations with the conservation laws of the matter fields (see Eqs.~(\ref{Ray1})-(\ref{gcont1}) in \S~\ref{ssEKFs}). This transforms the aforementioned set of differential formulae into the propagation equations of the three density parameters, namely
\begin{equation}
\dot{\Omega}_{\rho}= -\left[3(1+w)-2(1+q)\right]\,H\Omega_{\rho}\,, \hspace{15mm} \dot{\Omega}_{\psi}= -\left[3(1+w^{(\psi)})-2(1+q)\right]\,H\Omega_{\psi}  \label{dOmega1}
\end{equation}
and
\begin{equation}
\dot{\Omega}_K= 2qH\Omega_K\,,  \label{dOmega2}
\end{equation}
with $q$ given by (\ref{q1}). Replacing time with the dimensionless variable $\eta$, defined so that $\eta=\ln a$ and therefore giving ${\rm d}\eta/{\rm d}t=H$ \cite{UL}, the above system acquires the autonomous form
\begin{equation}
\Omega^{\prime}_{\rho}= -\left[3(1+w)-2(1+q)\right]\, \Omega_{\rho}\,, \hspace{15mm} \Omega^{\prime}_{\psi}= -\left[3(1+w^{(\psi)})-2(1+q)\right]\,\Omega_{\psi}  \label{auton1a}
\end{equation}
and
\begin{equation}
\Omega^{\prime}_K= 2q\Omega_K\,,  \label{auton1b}
\end{equation}
respectively.\footnote{In principle, one could express the set of (\ref{Ray1})-(\ref{gcont1}) into an alternative autonomous system of differential equations, by using the dynamical variables $x=\kappa\dot{\psi}/\sqrt{6}H$ and $y=\kappa\sqrt{V}/\sqrt{3}H$ proposed in~\cite{CLW}.} Also note that primes indicate differentiation with respect to $\eta$. Next, we will identify the equilibrium (fixed) points of the above system by setting $\Omega^{\prime}_{\rho}=0= \Omega^{\prime}_{\psi}=\Omega^{\prime}_K$, in which case Eqs.~(\ref{auton1a}) and (\ref{auton1b}) reduce to the constraints
\begin{equation}
\left[3(1+w)-2(1+q)\right]\,\Omega_{\rho}= 0\,, \hspace{15mm} \left[3(1+w^{(\psi)})-2(1+q)\right]\,\Omega_{\psi}= 0  \label{fixed1}
\end{equation}
and
\begin{equation}
q\Omega_K= 0\,.  \label{fixed2}
\end{equation}

\subsection{The fixed points}\label{ssFPs}
Demanding that $\Omega_{\rho}$, $\Omega_{\psi}\neq0$ always, we may distinguish between two alternative families of fixed points, namely those with $\Omega_K=0$ and those with $\Omega_K\neq0$. In the former case, Eqs.~(\ref{fixed1}) and (\ref{fixed2}) accepts equilibrium points of the form $(\Omega_{\rho},\,\Omega_{\psi},\,0)$ with
\begin{equation}
\Omega_{\rho}+ \Omega_{\psi}= 1 \hspace{10mm} {\rm and} \hspace{10mm} w=w^{(\psi)}= {1\over3}\,(2q-1)\,.  \label{fam1}
\end{equation}
When $\Omega_K\neq0$, on the other hand, the fixed points have $\Omega_{\rho}+\Omega_{\psi}+\Omega_K=1$ and $q=0$. The latter, together with constraints (\ref{fixed1}), ensures that
\begin{equation}
(1+3w)\Omega_{\rho}= 0 \hspace{10mm} {\rm and} \hspace{10mm} (1+3w^{(\psi)})\Omega_{\psi}= 0\,.  \label{fam2}
\end{equation}
which mean that $w=w^{(\psi)}=-1/3$ (given that $\Omega_{\rho}$, $\Omega_{\psi}\neq0$ always in our case).

It follows that the first family of equilibrium points corresponds to spatially flat FRW cosmologies, which decelerate when $w=w^{(\psi)}>-1/3$ and accelerate in the opposite case. The second group of fixed points, on the other hand, contains ``coasting'' universes (with $q=0$ and $a\propto t$) and nonzero spatial curvature. Note that the fact that $w=w^{(\psi)}$ in both equilibrium families means that, if one of the two matter components dominates over the other at a given instant, it will do so at all times. In the special case with $w=w^{(\psi)}=-1/3$, both matter fields have zero contribution to the effective gravitational mass of the system, which explains the ``coasting'' nature of the universal expansion. Finally, we should point out that the two aforementioned equilibrium states do not distinguish which matter field dominates over the other.

\subsection{Stability of the equilibrium states}\label{ssSESs}
Before proceeding to test the linear stability of the equilibrium points, we will first employ expression (\ref{q1}) to eliminate the deceleration parameter from the right-hand sides of Eqs.~(\ref{auton1a}a), (\ref{auton1a}b) and (\ref{auton1b}). In so doing, the aforementioned formulae recast as
\begin{equation}
\Omega^{\prime}_{\rho}= -(1+3w)\Omega_{\rho}+ (1+3w)\Omega^2_{\rho}+ (1+3w^{(\psi)})\Omega_{\psi}\Omega_{\rho}\,,  \label{auton2a}
\end{equation}
\begin{equation}
\Omega^{\prime}_{\psi}= -(1+3w^{(\psi)})\Omega_{\psi}+ (1+3w^{(\psi)})\Omega^2_{\psi}+ (1+3w^{(\psi)})\Omega_{\rho}\Omega_{\psi}  \label{auton2b}
\end{equation}
and
\begin{equation}
\Omega^{\prime}_K= (1+3w)\Omega_{\rho}\Omega_K+ (1+3w^{(\psi)})\Omega_{\psi}\Omega_K\,,  \label{auton2c}
\end{equation}
respectively. Let us now consider small homogeneous perturbations, around the fixed points (denoted by $(\bar{\Omega}_{\rho},\, \bar{\Omega}_{\psi},\,\bar{\Omega}_K)$ from now on), of the form
\begin{equation}
\Omega_{\rho}= \bar{\Omega}_{\rho}+ \omega_{\rho}\,, \hspace{10mm} \Omega_{\psi}= \bar{\Omega}_{\psi}+ \omega_{\psi} \hspace{10mm} {\rm and} \hspace{10mm} \Omega_K= \bar{\Omega}_K+ \omega_K\,,  \label{prts}
\end{equation}
where $|\omega|\ll|\bar{\Omega}|$ by default. Substituting the above into (\ref{auton2a})-(\ref{auton2c}) the latter transform into the (nonlinear) propagation equations of the $\omega$-perturbations. When linearised, these read
\begin{equation}
\omega^{\prime}_{\rho}= -\left[(1+3w)(1-2\bar{\Omega}_{\rho}) -(1+3w^{(\psi)})\bar{\Omega}_{\psi}\right]\omega_{\rho}+ (1+3w^{(\psi)})\bar{\Omega}_{\rho}\omega_{\psi}\,,  \label{omega1'}
\end{equation}
\begin{equation}
\omega^{\prime}_{\psi}= (1+3w)\bar{\Omega}_{\psi}\omega_{\rho}- \left[(1+3w^{(\psi)})(1-2\bar{\Omega}_{\psi}) -(1+3w)\bar{\Omega}_{\rho}\right] \omega_{\psi}  \label{omega2'}
\end{equation}
and
\begin{equation}
\omega^{\prime}_K= (1+3w)\bar{\Omega}_K\omega_{\rho}+ (1+3w^{(\psi)})\bar{\Omega}_K\omega_{\psi}+ \left[(1+3w)\bar{\Omega}_{\rho} +(1+3w^{(\psi)})\bar{\Omega}_{\psi}\right]\omega_K\,.  \label{omega3'}
\end{equation}
The above set of differential equations monitors the linear evolution of the perturbations and therefore the linear stability of the equilibrium states.

Let us now return to the first family of fixed points, which are characterised by the constraints $\bar{\Omega}_K=0$, $\bar{\Omega}_{\rho}+ \bar{\Omega}_{\psi}=1$ and $w=w^{(\psi)}=(2q-1)/3$. Then, written in matrix form, the system (\ref{omega1'})-(\ref{omega3'}) reads
\begin{eqnarray}
\left(\begin{matrix}
\omega^{\prime}_{\rho} \\ \omega^{\prime}_{\psi} \\ \omega^{\prime}_K
\end{matrix}\right)=
\left(\begin{matrix}
2q\bar{\Omega}_{\rho} & 2q\bar{\Omega}_{\rho} & 0 \\ 2q\bar{\Omega}_{\psi} & 2q\bar{\Omega}_{\psi} & 0 \\
0 & 0 & 2q
\end{matrix}\right)
\left(\begin{matrix}
\omega_{\rho} \\ \omega_{\psi} \\ \omega_K
\end{matrix}\right)\,,  \label{matrix1}
\end{eqnarray}
keeping in mind that $2q=1+3w=1+3w^{(\psi)}$. The characteristic polynomial of the $3\times3$ matrix seen above is $\lambda(\lambda-2q)^2=0$, with eigenvalues $\lambda_1=0$ and $\lambda_2=\lambda_3=2q$. Given that the presence of a zero eigenvalue complicates the standard stability analysis considerably (e.g.~see~\cite{BS1,BS2} for an extensive discussion of the issue), we will seek analytic solutions. These can be obtained by assuming, for example, that the coefficients $\bar{\Omega}_{\rho}$, $\bar{\Omega}_{\psi}$ and $q$ are all constant. In practice, this means considering a period in the model's evolution, during which its deceleration/accleration rate is constant and the energy-density contributions of both matter fields remain unchanged. In such a case, the system (\ref{matrix1}) solves to give
\begin{equation}
\omega_{\rho}= \mathcal{C}_1\left(\bar{\Omega}_{\psi} +\bar{\Omega}_{\rho}{\rm e}^{2qt}\right)+ \mathcal{C}_2\bar{\Omega}_{\rho}\left(1-{\rm e}^{2qt}\right)\,,  \label{family1a}
\end{equation}
\begin{equation}
\omega_{\psi}= \mathcal{C}_3\bar{\Omega}_{\psi}\left(1-{\rm e}^{2qt}\right)+ \mathcal{C}_4\left(\bar{\Omega}_{\rho}+ \bar{\Omega}_{\psi}{\rm e}^{2qt}\right)  \label{family2b}
\end{equation}
and
\begin{equation}
\omega_K= \mathcal{C}_5\,{\rm e}^{2qt}\,,  \label{family2c}
\end{equation}
with $\mathcal{C}_1,\dots,\mathcal{C}_5$ being the integration constants. Clearly, the late-time behaviour of the above solution depends on the sigh of the deceleration parameter. More specifically, when $q$ is positive, all three perturbations diverge. In the opposite case, on the other hand, $\omega_K\rightarrow0$, while the other two tend to constant values (with $\omega_{\rho}\rightarrow\mathcal{C}_1\bar{\Omega}_{\psi}+ \mathcal{C}_2\bar{\Omega}_{\rho}$ and $\omega_{\psi}\rightarrow\mathcal{C}_3\bar{\Omega}_{\psi}+ \mathcal{C}_4\bar{\Omega}_{\rho}$). Given that $2q=1+3w=1+3w^{(\psi)}$, the stability of the fixed points is determined by the equation of state of the matter fields involved. When the latter satisfies the strong energy condition, we have unstable repellers. In the opposite case, the associated equilibrium states may be seen as stable attractors, though generally only in the Lyapunov sense.\footnote{It is conceivable that, by selecting ``favourable'' initial conditions, both $\omega_{\rho}$ and $\omega_{\psi}$ might tend to zero at late tines. Nevertheless, even if it so happens, it would be the exception to the rule.}

The second family of equilibrium states has $\bar{\Omega}_K\neq0$, $q=0$ and $w=w^{(\psi)}=-1/3$. These severely constrain the linear evolution of the $\omega$-perturbations (see Eqs.~(\ref{omega1'})-(\ref{omega3'}) above), the propagation formulae of which reduce to
\begin{equation}
\omega^{\prime}_{\rho}= 0\,, \hspace{10mm} \omega^{\prime}_{\psi}= 0 \hspace{10mm} {\rm and} \hspace{10mm} \omega^{\prime}_K= 0\,,  \label{family2}
\end{equation}
respectively. Consequently, at the linear level, all three perturbations remain constant, thus making the associated family of fixed points stable \`{a} la Lyapunov.

Before closing our discussion on the dynamical stability of the equilibrium states, we should remind the reader that none of the fixed points identified in this section corresponds to the critical static case presented in \S~\ref{sFUGRM} earlier. Both families of equilibrium states are non-static cosmologies, which undergo accelerated, decelerated, or coasting expansion. The stability of static ghost-like FRW cosmologies, against general inhomogeneous perturbations, has been studied in~\cite{BT} by means of relativistic cosmological perturbation theory. There it was found that linear inhomogeneities in the density distribution of both matter fields remain constant in time, which implies Lyapunov-type stability against this (scalar) type of distortions. Neutral stability, namely oscillatory with constant amplitude, was also found in the case of linear (pure tensor) gravitational-wave perturbations. That behaviour appears be in general agreement with the stability results reported here.

\section{Discussion}\label{sD}
Cosmological scenarios that allow for ghost-like matter are becoming more frequent in the cosmological literature, though they are still far from mainstream. Ghost fields are attractive, since they can offer phenomenological alternatives to questions as fundamental as inflation and the nature of dark energy. Perhaps the most typical example is the fairly popular Quintom scenario, which combines two scalar fields with canonical and negative kinetic terms respectively. On the other hand, ghosts have been reported to suffer from quantum-level instabilities, which could impose strict constraints on their presence. Nevertheless, the debate around the ghost/quintom hypothesis is still going on and the issue may only be settled by a successful theory of quantum gravity. Technically speaking, there are more than one ways of realising ghost matter. In this study, we did so by assuming a specific Lagrangian with an energy-momentum tensor that corresponds to a perfect fluid with negative (effective) energy density. The specifics of the aforementioned ghost component are such that it never enters a tachyon-like, or a phantom-like, regime. We then looked into the kinematical and dynamical phenomenology of FRW models containing conventional matter and an always subdominant, non-interacting ghost component. Therefore, to some extent, our scenario resembles that of the aforementioned Quintom model. When introduced into a Friedmann universe, this type of matter has been known to play a role that phenomenologically resembles that of positive spatial curvature. In this work we have taken another look at this resemblance and its potential implications.

After a brief discussion on the technicalities of ghost-like scalar fields and their fluid description, we considered an FRW model filled with a non-interacting mixture of ordinary and ghost-like matter. The opposing contribution of the two aforementioned sources to the total energy density of the model means that there can be a static moment during the model's lifetime, even in the absence of positive curvature. Whether such a critical stage occurs in the future, or in the past, depends on the evolution of the ghost component relative to the ordinary matter. In either case, the static moment is followed by a bounce. Put a another way, an expanding universe cames to a halt and then starts to collapse, whereas a contracting model re-bounces at a finite ``radius'' (before it reaches a singularity) and then begins to expand. What is most intriguing, is that there is a relatively simple condition for this cyclic pattern of expansion, collapse and re-expansion to repeat itself indefinitely. Moreover, at no time the ghost matter needs to dominate over its conventional counterpart, which means that the energy density of the total matter remains non-negative. In fact, the ghost component is assumed to be completely subdominant at present.

The linear stability of analogous static ghost-cosmologies was investigated in~\cite{BT} using relativistic cosmological perturbation theory. That study allowed for inhomogeneous scalar and pure-tensor distortions, namely for density and gravitational-wave perturbations. The results showed Lyapunov-like stability against both types of distortions, since the perturbations were found to either remain constant in time, or oscillate with constant amplitude. Here, we employed dynamical-system techniques to test the stability of FRW comsologies with a mixture of ordinary and ghost matter, against homogeneous perturbations. Our analysis led to two families of (non-static) fixed points, distinguished by the absence or the presence of spatial curvature. Unfortunately, both families were rather severely constrained and therefore fairly limited phenomenologically. In the absence of curvature, we found that the linear stability of the family members depended on the equation of state of the matter fields involved. When the latter satisfied the strong energy condition, the associated fixed points were unstable repellers, while in the opposite case they were found to be stable in the Lyapunov sense. In the case of nonzero spatial curvature, the stability question was more straightforward to answer, since all the related fixed points were stable \`a la Lyapunov coasting cosmologies.\\

\textbf{Acknowledgments:} We would like to thank Martin Richarte, Peter Stichel and David Vasak for their helpful comments. AC wishes to thank the Astronomy Lab at the Aristotle University of Thessaloniki, where the initial part of this work was done, for their hospitality. JDB was supported by the
Science, Technology and Facilities Council (STFC) of the UK. CGT acknowledges support by the Hellenic Foundation for Research and Innovation (H.F.R.I.), under the “First Call for H.F.R.I. Research Projects to support Faculty members and Researchers and the procurement of high-cost research equipment grant” (Project Number: 789).

\end{document}